\documentclass[prb,twocolumn,showpacs]{revtex4}
\usepackage{graphicx}
\usepackage{amsmath}
\usepackage{amssymb}
\usepackage{dcolumn}
\usepackage{float}
\usepackage{bm}
\usepackage{multirow}
\usepackage{booktabs}
\usepackage{xcolor}
\newcolumntype{M}[1]{>{\centering\arraybackslash}m{#1}}

\newcommand{\nn}{\nonumber}
\renewcommand{\Re}{\mathrm{Re}\,}

\renewcommand{\Re}{\mathrm{Re}\,}

\newcommand{\bs}{\boldsymbol}
\DeclareMathAlphabet{\bi}{OML}{cmm}{b}{it}

\def\be{\begin{equation}}
\def\ee{\end{equation}}
\def\bearr{\begin{eqnarray}}
\def\eearr{\end{eqnarray}}
\def\la{\langle}
\def\ra{\rangle}

\usepackage[title,titletoc,toc]{appendix}

\begin{document}
\title{Drude weight and optical conductivity of a two-dimensional heavy-hole gas with $k$-cubic 
spin-orbit interactions}
\bigskip
\author{Alestin Mawrie and Tarun Kanti Ghosh\\
\normalsize
Department of Physics, Indian Institute of Technology-Kanpur,
Kanpur-208 016, India}
\date{\today}
 
\begin{abstract}
We present detailed theoretical study on zero-frequency Drude weight and optical 
conductivity of a two-dimensional heavy-hole gas(2DHG) with $k$-cubic Rashba and 
Dresselhaus spin-orbit interactions. 
The presence of $k$-cubic spin-orbit 
couplings strongly modifies the Drude weight in comparison to the electron gas
with $k$-linear spin-orbit couplings. 
For large hole density and strong $k$-cubic spin-orbit
couplings, the density dependence of Drude weight deviates from the linear behavior.
We establish a relation between optical conductivity and 
the Berry connection. Unlike two-dimensional electron gas with $k$-linear spin-orbit 
couplings, we explicitly show that the optical conductivity does not vanish even 
for equal strength of the two spin-orbit couplings. We attribute this fact to the 
non-zero Berry phase for equal strength of $k$-cubic spin-orbit couplings.
The least photon energy needed to set in the optical transition
in hole gas is one order of magnitude smaller than that of electron gas.
Types of two van Hove singularities appear in the optical  
spectrum are also discussed.

\end{abstract}

\maketitle
\section{introduction}
Spin-orbit coupling \cite{soc,soc1,soc2,soc3} plays a vital role in several physical 
properties of various systems because it breaks the spin degeneracy even 
at zero magnetic field. There are mainly two kinds of spin-orbit interaction in condensed
matter systems, namely Rashba spin-orbit coupling due to inversion
asymmetry of the confining potential \cite{Rashba1,Rashba2} and Dresselhaus spin-orbit coupling 
generated by the asymmetry of the host bulk crystals \cite{luttinger1,dres,luttinger2}. 
The Rashba spin-orbit coupling has been realized in 
various systems such as zincblende semiconductor quantum wells \cite{rashba-exp1}, 
carbon nanotubes \cite{cnt}, two-dimensional materials \cite{graphene-soc1,graphene-soc2},  
and neutral atomic Bose-Einstein condensates \cite{bec1,bec2}.
The spin-orbit interaction (SOI) is essential in controlling the spin degree of 
freedom of the charge carrier in spin-based devices. 
In most of the studies systems with $k$-linear Rashba and Dresselhaus 
spin-orbit interactions(RSOI and DSOI) are of much interest. 
However higher order momentum-dependent spin-orbit interactions 
have also been seen to dominate in many physical systems.
For example, $k$-cubic RSOI dominates in 
two-dimensional heavy-hole gas formed at the p-type {\mbox GaAs/AlGaAs} 
heterostructure \cite{sherman,winkler,winkler1},
in two-dimensional electron gas formed at the surface of the inversion symmetric 
oxide {\mbox SrTiO}${}_3$ \cite{SrTiO3} and in strained {\mbox Ge/SiGe} 
quantum wells \cite{strain}.

The optical conductivity is due to the transitions from one energy level to 
another energy level, whereas the zero-frequency Drude weight is associated
with the intra-level transitions. 
The real part of the complex frequency-dependent longitudinal optical
conductivity provides the absorption as a function of photon energy.
Its measurement through optical spectroscopy is an important tool for 
probing shape of the spin-split energy levels. 
Typical order of the spin-split energy is the same as that of an electromagnetic
radiation with terahertz (THz) frequency. The high-frequency radiation  
plays an important role to control the spinors of the spin-split energy 
levels due to the SOIs. 
Optical response studies will be useful for the high-speed electronic devices 
since the radiation with THz frequency flips the spin in a very short time.
It opens the possibility of seeing the resonance effects through the 
optical transition between spin-split levels and leads to unique spectral features.

Several theoretical studies of the optical conductivity have been 
carried out on spin-orbit coupled electron systems formed at the
semiconductor heterojunctions \cite{e-op1,e-op2} as well as in 
$t_{2g}$ bands of an oxide with perovskite structure \cite{mac}.
The optical spectrum of hole gas
with $k$-cubic RSOI has been studied \cite{hole-op} partially. 
The connection between optical longitudinal charge conductivity and optical 
spin Hall conductivity in electron and hole systems have been established 
in Refs. \cite{e-op2,shc}.
In Ref.  \cite{nature-phys},  it is shown that the optical conductivity disappears when
the linear DSOI is same as the RSOI of two-dimensional electron systems.
This optical conductivity has also been studied in various
single layer two-dimensional materials like graphene \cite{graphene-opt}, 
MoS$ {}_2 $ \cite{mos2-opt}, silicene \cite{silicene-opt} and 
surface states of topological insulators \cite{ti-opt}.

In this work we present zero-frequency Drude weight and optical conductivity of 
a two-dimensional heavy-hole gas with $k$-cubic RSOI and DSOI. 
The Drude weight is strongly modified due to the presence of the $k$-cubic 
spin-orbit couplings. We obtain an analytical expression of the Drude
weight when only the $k$-cubic RSOI is present. It deviates from the linear 
density dependence for large hole
density and for strong spin-orbit couplings. It decreases with the increase of 
the spin-orbit couplings. 
We show that the optical conductivity is directly related to the Berry connection and
does not vanish even for equal strength of the two spin-orbit couplings.
The minimum photon energy required for the onset of the optical transition 
in hole gas is one order of magnitude smaller than that of electron gas.
We also identify the nature of the two van Hove singularities appear in the optical 
spectrum.

We organise this paper as follows.
In section II, we present basic ground state properties of the heavy-hole system with
the $k$-cubic RSOI and DSOI. 
In section III, we present the effect of spin-orbit interactions on the 
Drude weight and study various aspects of the optical conductivity for the same
system at zero as well as non-zero temperature. Types of the van Hove singularities
are tabulated in this section.  
A summary of our main results are provided in section IV.

\section{Physical system}
The Hamiltonian of a heavy-hole with the both $k$-cubic RSOI and DSOI is given by \cite{Liu,Zarea,Denis}
\begin{eqnarray}\label{hamil}
H & = & \frac{{\bf p}^2}{2m^\ast} + 
\frac{i\alpha}{2\hbar^3}\big(p_{-}^3\sigma_{+}-p_{+}^3\sigma_{-}\big)\nonumber\\
&-&\frac{\beta}{2\hbar^3}\big(p_{-}p_+p_-\sigma_{+}+p_{+}p_{-}p_+\sigma_{-}\big),
\end{eqnarray}
where $m^\ast$ is the effective mass of the hole,
$p_\pm=p_x \pm ip_y$ and $\sigma_\pm=\sigma_x\pm i\sigma_y$,
with $\sigma_x$ and $\sigma_y$ are the Pauli's matrices. 
Also, $\alpha$ is the strength of RSOI 
which measures the structure inversion asymmetry-induced splitting  
and $\beta$ is the strength of DSOI which measures the
bulk inversion asymmetry-induced spin splitting in the system.
Typical value of Rashba strength in narrow gap semiconductor is 
$\alpha \sim 10^{-22} $ eV-cm${}^3$ and $ \beta$ is always less than $\alpha$.

The energy spectrum and the corresponding eigenstates are given by  
\begin{eqnarray}\label{NMa}
E_{\lambda}({\bf k}) = \frac{\hbar^2k^2}{2m^\ast}+\lambda k^3
\Delta(\theta)
\end{eqnarray}
and 
\begin{eqnarray}
\Psi_{\bf k}^\lambda ({\bf r}) = \frac{\exp( i{\bf k} \cdot {\bf r}) }{\sqrt{2\Omega}}
\begin{pmatrix}
1\\ \lambda e^{i (2\theta - \phi)}
\end{pmatrix},
\end{eqnarray}
where $\lambda = \pm $ denotes spin-split energy levels, 
$\Omega $ is the surface area of the two-dimensional system,
and $ \Delta (\theta) = \sqrt{\alpha^2+\beta^2-2\alpha\beta \sin2\theta}$
is the angular anisotropic term with
$ \theta = \tan^{-1}(k_y/k_x)$ and 
$\phi = \tan^{-1}(\alpha \cos\theta - \beta \sin\theta)/(\alpha \sin\theta - \beta  \cos\theta)$.  
The presence of both the SOIs is responsible for the 
angular anisotropy of the energy spectra. 
The spin splitting energy 
$ E_g({\bf k}) = E_+({\bf k}) - E_-({\bf k}) = 2 k^3 \Delta(\theta) $ 
between two branches is also anisotropic. The maximum and minimum spin splitting
occurs at $ \theta = 3\pi/4$ or $7\pi/4$  and $ \theta = \pi/4$ or $ 5\pi/4$; 
and the corresponding spin splitting energy values are 
$ E_g = 2 k^3 (\alpha \pm \beta)$, respectively.

The Berry connection for this system is given by  
\begin{eqnarray}\label{BerryC}
{\bf A_k} & = & i \la \psi_{{\bf k}}^{\lambda}
\vert  {\bs \nabla}_{\bf k} \vert \psi_{{\bf k}}^{\lambda} \ra =
\bigg(\frac{3 \alpha^2 + \beta^2  - 4\alpha \beta \sin 2\theta}{\Delta^2(\theta)}\bigg)
\frac{\hat \theta}{k},
\end{eqnarray}
where $\hat \theta = -\sin \theta \, \hat x + \cos \theta \, \hat y$ is the unit polar vector.
The corresponding Berry phase is given as
\begin{eqnarray}\label{BerryP}
\gamma & = &  \oint {\bf A_k}\cdot {\bf dk} = 
\pi+ 2\pi\frac{\alpha^2-\beta^2}{\vert \alpha^2-\beta^2\vert}.
\end{eqnarray}
The Berry connection and Berry phase do not vanish for $\alpha=\beta $ case,
in complete contrast to the electron gas with equal strength of the spin-orbit 
couplings \citep{nature-phys,ti-opt}.

First we shall calculate density of states (DOS) of the spin-split energy branches,
required for the calculation of the Fermi energy ($E_f$) and the associated anisotropic
Fermi wave vectors $k_f^{\lambda}(\theta)$.
The density of states of the spin-split energy branches are obtained from
\begin{eqnarray}
D_{\lambda}(E) & = & \frac{1}{(2\pi)^2} \int d^2k  
\delta(E-E_{\lambda}({\bf k} ) ) \nn \\
& = &  \frac{D_0}{2\pi} \int d^2k
\frac{\delta(k - k_{E}^{\lambda}(\theta))}{|k_{E}^{\lambda}(\theta) + 
\lambda 6 \pi D_0 \Delta (\theta) ( k_{E}^{\lambda}(\theta) )^2 |},
\end{eqnarray}
where $D_0= m^*/(2\pi \hbar^2)$ is the DOS of spin-polarized hole gas
without SOIs and $ k_{E}^{\lambda} $ is the real solution of
the cubic equation 
$(\hbar k_{E}^{\lambda})^2/2m^\ast + \lambda \Delta(\theta) (k_E^{\lambda})^3 - E = 0$.
The energy dependence of the density of states is shown in Fig. 1
for two different values of SOIs.
The DOS of the heavy-hole spin-split energy branches change asymmetrically 
with respect to $D_0$, whereas it changes symmetrically with respect to $D_0$ for
$k$-linear spin-orbit coupled electron systems \cite{soc}.
\begin{figure}[!htbp]
\begin{center}\leavevmode
\includegraphics[width=95mm,height=65mm]{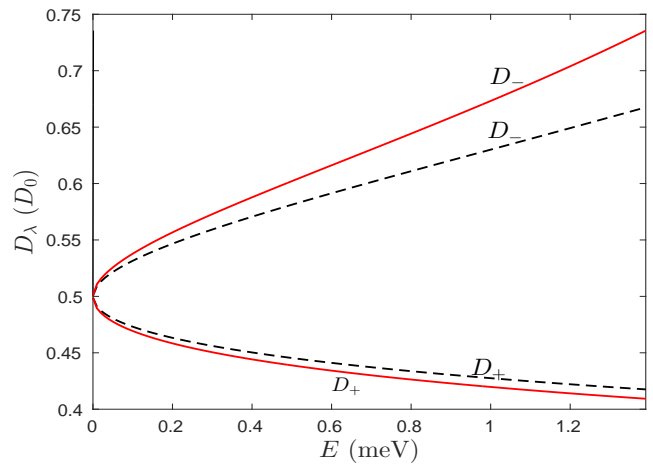}
\caption{(color online) Plots of $ D_{\pm}(E) $
(in units of $ D_0$) vs energy $E$ for $\alpha = 0.08$ eV nm$^3$ (dashed) 
and $ \alpha =0.1 $ eV nm$^3$ (solid) with $\beta=0.6\alpha$.}
\label{Fig1}
\end{center}
\end{figure}

For a given set of system parameters 
($n_h$ being the heavy-hole density, $\alpha$ and $\beta$), the Fermi
energy $E_f$ can be evaluated numerically from the normalization condition:
$ n_h = \sum_{\lambda}\int_0^{E_f} D_{\lambda}(E) dE $.

For $\beta =0 $, the exact analytical expressions of the Fermi wave vectors 
\cite{loss} are given by
$
k_{f}^{0,\pm} = \sqrt{3\pi n_h - L/(8 l_{\alpha}^2) } \mp 
L/(4 l_{\alpha})$,
where $ L = (1- \sqrt{1-16 \pi n_h l_{\alpha}^2 }) $ and $l_{\alpha} = m^* \alpha/\hbar^2$.
When $\beta \neq  0$, it is not possible, due to the anisotropic nature of the
spectrum, to derive exact analytical expressions
of the spin-split anisotropic Fermi wave vectors $k_f^{\lambda}(\theta)$. Therefore, we
numerically calculate $k_f^{\lambda}(\theta)$ from the following cubic equation:
$ (\hbar k)^2/2m^* + \lambda k^3 \Delta(\theta) - E_f = 0 $,
for given values of $\alpha$, $\beta$, $n_h$ and $E_f$.
The anisotropic Fermi contours $k_f^{\lambda}(\theta)$, shown in Fig. 6, are symmetric 
with respect to the lines $ k_y \pm k_x =0 $.

\section{Drude weight and Optical Conductivity}
Consider a two-band system of charge carriers, electron/hole, subjected to an 
oscillating electric field (${\bf E} \sim \hat {\bf x} E_0 e^{i\omega t} $).
The complex charge current conductivity is given by
$$
\Sigma_{xx}(\omega) = \sigma_D(\omega) + \sigma_{xx}(\omega),
$$
where $ \sigma_D(\omega) = \sigma_d/(1-i \omega \tau) $ is the dynamic 
Drude conductivity due to the intra-band transitions, with 
$\sigma_d $ being the static Drude conductivity and 
$\sigma_{xx}(\omega) $ is the complex optical conductivity due to the inter-band optical
transitions between two branches. Also, $\tau $ is the momentum relaxation
time. 
It is to be noted here that real part of $\sigma_D $ and $\sigma_{xx} $
correspond to the absorptive parts of the optical transition.
It implies that absorption peaks in real parts of the conductivities will display dips
in the experimentally measured transmission.

The real part of the Drude conductivity is 
$ {\rm Re} \, \sigma_D(\omega) = D_w \delta(\omega) $, 
where $ D_w = \pi \sigma_d/\tau $ is known as Drude weight.
It shows that the peak appears around $\omega = 0 $. 
On the other hand, $ {\rm Re} \, \sigma_{xx}(\omega) $ is  
a function of photon energy with vanishing momentum
($q \rightarrow 0$).
Here the vanishing momentum limit displays the fact that 
the momentum of the charge carrier is not altered by the
electron-photon interaction.
An optical absorption occurs through inter-band transition from
$\lambda = - $ branch to $\lambda = + $ branch
and helps to make spin-flip transition from one spin branch to 
another spin branch.

\subsection{Drude weight}
Using the semi-classical Boltzmann transport theory \cite{mermin}, the Drude weight at 
very low temperature can be written as
\bearr \label{drude-weight}
D_w & = & \frac{e^2 }{4 \pi} \sum_{\lambda} \int d^2k \la \hat v_{x}\ra^2_{\lambda} 
\delta(E_{\lambda}({\bf k}) - E_f),
\eearr
where $\hat v_x$ is the $x$-component of the velocity operator and $E_f$ is the Fermi energy 
for a given system.
Using Eq. (\ref{drude-weight}), we have calculated (see Appendix for detail calculation) 
the Drude weight ($D_{w}^e$) of a two-dimensional electron gas with $k$-linear RSOI, and it is given by
\be
D_{w}^e = \frac{\pi e^2}{m_e} \Big( n_e - \frac{\alpha_e^2 m_{e}^2}{2\pi \hbar^4}\Big).
\ee
Here $m_e $ is the effective mass of an electron, $\alpha_e$ is the spin-orbit
coupling strength and $n_e $ is the density of electrons.
This result exactly matches with the result obtained in 
Ref. \cite{agarwal}.

For the present problem, the $x$-component of the velocity operator is given by 
\bearr
\hat{v}_x & = & v_x I + V_1 \sigma_x - V_2 \sigma_y \label{vop1}
\eearr
where $v_{x}= v \cos \theta$,
$V_1= 3v_{\alpha} \sin 2\theta-v_\beta(2+\cos 2\theta) $,
$ V_2 = 3v_{\alpha} \cos 2\theta + v_\beta\sin 2\theta$,
with $v = \hbar k/m $, $ v_{\alpha} = \alpha k^2/\hbar $ and
$ v_{\beta} = \beta k^2/\hbar $.
Also, $ \la \hat v_x \ra_{\lambda} $ is the average value of 
the operator $\hat v_x$ with respect to the state $\psi_{\bf k}^{\lambda}({\bf r}) $.
After simplification, it reduces to
\bearr
D_w & = & \frac{e^2 }{4 \pi} \sum_{\lambda} \int d\theta
\frac{B_{\lambda}({\bf k})}{|\hbar^2 k_f^{\lambda}/m^* 
+ \lambda 3 \Delta(\theta) (k_f^{\lambda})^2|},  
\eearr
where
$ B_{\lambda}({\bf k}) = [(\hbar k_f^{\lambda}/m^*) \cos \theta
+ \lambda ((k_f^{\lambda})^2/\hbar) \{3 \alpha \sin \phi - \beta \cos \phi
- 2 \beta \cos(2\theta - \phi)\}]^2.
$
For $\beta=0$, we have the following analytical expression of $D_w$:
\bearr \label{Drude-alpha}
D_{w}^{R} = \frac{3e^2}{16m^* l_{\alpha}^2} \Big[1 - \frac{56}{3}\pi n_h l_{\alpha}^2
- (1-16 \pi n_h l_{\alpha}^2)^{3/2}\Big].
\eearr
For $\alpha =0 $ and $\beta \neq 0$, the Drude weight can be obtained from 
Eq. (\ref{Drude-alpha}) by replacing $l_{\alpha} $ by $l_{\beta}= m^*\beta/\hbar^2 $.
The variations of $D_w $ with $n_h $ and $\alpha $ are shown in Fig. 2.
It is known that the Drude weight varies linearly with the carrier density 
for a free fermion as well as for 2DEG with $k$-linear spin-orbit couplings 
\cite{mermin,agarwal,drude}.
Equation (\ref{Drude-alpha}) and the left panel of Fig. 2 clearly show deviation 
from the linear density dependence of $D_w$ for large density and strong 
RSOI.
On the other hand, the Drude weight decreases with the increase of the spin-orbit
couplings as shown in the right panel of Fig. 2. 

\begin{figure}[!htbp]
\begin{center}\leavevmode
\includegraphics[width=88mm,height=65mm]{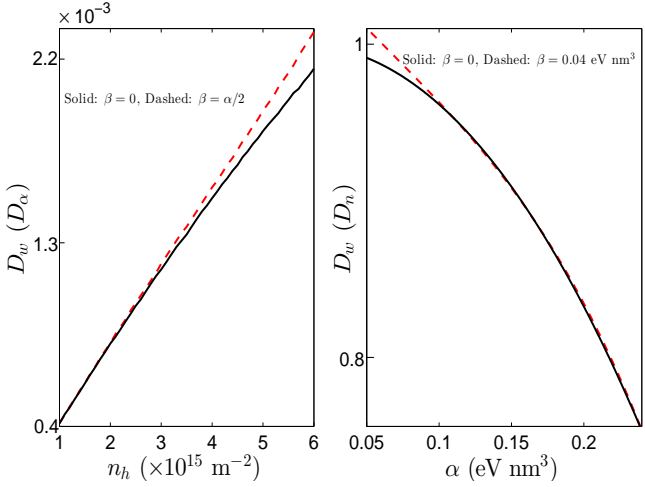}
\caption{(color online) Left panel: Plots of $D_w$ (in units of 
$D_{\alpha} = e^2\pi/m^*l_{\alpha}^2$)
vs $n_h$ for $\alpha = 0.12 $ eV nm${}^3$.
Right panel: Plots of $D_w$ (in units of $D_n= e^2n_h\pi/m^*$) vs $\alpha $ 
for $n_h = 2.0\times 10^{15}$ m${}^{-2}$.}
\label{Fig1}
\end{center}
\end{figure}

It should be mentioned here that the effect of the Coulomb interaction 
is not taken into account in the above discussion. There may or may not 
be a significant effect of the Coulomb interaction on $D_w$ when the 
interaction parameter $r_s$ is very large (i.e. at the very low densities 
$n_h < 0.5 \times 10^{15} $ m${}^{-2}$). For example, the strong effect of 
the Coulomb interaction in low-density 2DHG leads to the negative compressibility 
observed in Ref. \cite{compressibility}. On the other hand, based on electron 
measurements \cite{e-spin}, one would expect an enhanced spin susceptibility at 
very low densities for 2DHG but it is not observed experimentally
\cite{anomalous}. The behavior of electrons and holes can be quite different, 
due to different effective mass and different form of spin-orbit coupling, 
as revealed in several theoretical studies \cite{coulomb,coulomb1,coulomb2}.

\subsection{Optical conductivity} 
The Kubo formula for the $xx$ component of the 
optical conductivity in terms of the Matsubara Green's function is given by
\begin{eqnarray}\label{NMac}
& &\sigma_{xx}(\omega)=  -\frac{e^2}{i\omega}\frac{1}{(2\pi)^2} 
\int_{0}^{\infty} \int_{0}^{2\pi} k dk d\theta \nonumber\\
& \times & T \sum_l {\rm Tr}\langle  \hat v_x\hat{G}({\bf k},\omega_l) \hat v_x 
\hat{G}({\bf k},\omega_s+\omega_l) \rangle_{i\omega_s\rightarrow \omega + i\delta }.
\end{eqnarray}
Here $T$ is the temperature,  $\omega_s=(2s+1)\pi T$ and $\omega_l=2l\pi T$ 
are the fermion and boson Matsubara frequencies, respectively, with $s$ and $l$ are integers.

The matrix Green's function for the two-level system associated
with the Hamiltonian (\ref{hamil}) is given by
\begin{eqnarray}
\hat{G}({\bf k},i\omega_s) & = & \Big[i\hbar\omega_s + \mu - (\hbar k)^2/2m^* - 
S_1 \sigma_x - S_2\sigma_y\Big]^{-1}
\nonumber\\
& = & \frac{i\hbar\omega_s + \mu - \frac{\hbar^2k^2}{2m^\ast} + S_1\sigma_x + S_2\sigma_y}
{(i\hbar\omega_s + \mu - \frac{\hbar^2k^2}{2m^\ast})^2 - S_1^2 - S_2^2}
\end{eqnarray}
with $S_1=k^3(-\alpha \sin3\theta+\beta\cos\theta)$
and $S_2=k^3(\alpha \cos3\theta+\beta\sin\theta)$.
It is convenient to write the Green's function as follows
\begin{eqnarray}
\hat{G}({\bf k},i\omega_s)=\frac{1}{2} \sum_\lambda \Big[ I + \lambda
{\bf F} \cdot {\bs \sigma} \Big] G_0({\bf k},\lambda,\omega_s),
\end{eqnarray}
where ${\bf F}=(S_1,S_2)/\sqrt{S_1^2+S_2^2}$,
and $G_0({\bf k},\lambda,\omega_s)=1/(i\hbar\omega_s+\mu-{\hbar^2k^2}/{2m^\ast} -
\lambda \sqrt{S_1^2+S_2^2})$.

Now we can write down $\hat{v}_x \hat{G}({\bf k},\omega_l)$ as below
\begin{eqnarray}
\hat{v}_x \hat{G}({\bf k},\omega_l)=\frac{1}{2}\sum_\lambda M_\lambda G_0({\bf k},\lambda,\omega_l),
\end{eqnarray}
 where $ M_\lambda = (v + V_1\sigma_x - V_2 \sigma_y)
(I+\lambda  {\bf F } \cdot {\bs \sigma} )$,
 which will give us
 \begin{eqnarray*}
& & {\rm Tr}\langle \hat v_x\hat{G}({\bf k},\omega_l) \hat v_x 
\hat{G}({\bf k},\omega_s+\omega_l) \rangle = \nonumber\\
& {} & \frac{1}{2}\sum_{\lambda\lambda^\prime}\Big[(1+\lambda\lambda^\prime )v^2 + 
V_1^2 + V_2^2 +\lambda \lambda^\prime \{(V_1F_x-V_2F_y)^2 \nn \\
& {} & -(V_1F_y+V_2F_x)^2 \} +  2(\lambda+\lambda^\prime)v(V_1F_x-V_2F_y)\Big] \nn \\  
& {} & \times  G_0({\bf k},\lambda,\omega_l)G_0({\bf k},\lambda^\prime,\omega_{s+l}).
\end{eqnarray*}

Using the identity
\begin{eqnarray}
& & T\sum_l\Big[\frac{1}{(i\hbar\omega_l+\mu-E_\lambda)} \cdot 
\frac{1}{i\hbar(\omega_s+\omega_l)+\mu-E_{\lambda^\prime}}  \Big]\nonumber \\ 
& = & \begin{cases}\frac{f(E_\lambda)-f(E_{\lambda^\prime})}
{i\hbar\omega_s-E_{\lambda^\prime}+E_\lambda}  &\text{$\lambda\neq\lambda^\prime$}\\
0 &\text{$\lambda=\lambda^\prime$}
\end{cases}
\end{eqnarray}
with the Fermi-Dirac distribution function 
$f(E)=[e^{(E-\mu)/(k_BT)}+1]^{-1}$,
one can see that the contribution of the intraband transition
($\lambda = \lambda^{\prime} $) to the optical conductivity is simply zero. This happens
as a result of the momentum conservation.
The non-zero contribution is coming only from the interband transitions 
($\lambda  \neq \lambda^{\prime})$. 
With this we can simplify further as
\begin{eqnarray*}
& & T\sum_l{\rm Tr}\langle \hat v_x\hat{G}({\bf k},\omega_l) \hat v_x 
\hat{G}({\bf k},\omega_s+\omega_l) \rangle \nonumber \\ & = & 
\frac{(S_2V_1+S_1V_2)^2}{S_1^2+S_2^2} 
\Big[ \frac{f(E_-)-f(E_+)}{i\hbar\omega_s-E_++E_-}+(E_-\leftrightarrow E_+) \Big],
\end{eqnarray*}
thereby we obtain the expression for the optical conductivity as follows
\begin{eqnarray}
\sigma_{xx}(\omega) & = -& \frac{e^2}{i(2\pi \hbar)^2\omega}
\int_{0}^{\infty} \int_{0}^{2\pi} k^5 \, H(\theta) dk d\theta \nonumber\\
& \times &\Big[\frac{f(E_-)-f(E_+)}{\hbar\omega + i\delta - E_+ + E_-} +
(E_-\leftrightarrow E_+)\Big],
\end{eqnarray}
where the explicit expression of the optical matrix element $H(\theta)$
is given by $ H(\theta) = \sin^2\theta 
\Big(3\alpha^2+\beta^2-4\alpha\beta\sin2\theta\Big)^2 /\Delta^2(\theta)$.
It is interesting to note that the above equation can be re-written in terms 
of the $x$-component of the Berry connection ($A_{k_x}$), which is given by
\begin{eqnarray}
\sigma_{xx}(\omega) & = & -\frac{e^2}{i(2\pi \hbar)^2\omega}
\int_{0}^{\infty} \int_{0}^{2\pi} k^7 \, \Delta^2(\theta) A_{k_x}^2 dk d\theta \nonumber\\
& \times &\Big[\frac{f(E_-)-f(E_+)}{\hbar\omega + i\delta - E_+ + E_-} +
(E_-\leftrightarrow E_+)\Big].
\end{eqnarray}
Similar connection has been established for MoS${}_2$ system \cite{mos2-opt}.

We have also calculated other components of the optical conductivity i.e.
$\sigma_{yy} (\omega) $ and $ \sigma_{xy}(\omega) $. We find that
$\sigma_{yy}(\omega) = \sigma_{xx} (\omega)$
and $\sigma_{xy}(\omega) = 0 = \sigma_{yx}(\omega) $.
It implies that the anisotropic Fermi contours do not lead to anisotropic
optical conductivity.
This is similar to the isotropic charge conductivity of a two-dimensional electron gas 
with combined RSOI and DSOI \cite{drude}.
Moreover, $ \sigma_{yy}(\omega) $ can also be expressed in terms of the
$y$-component of the Berry connection, similar to the $\sigma_{xx}(\omega) $ case.

Keeping in mind that $\omega >0 $
the absorptive part of the optical conductivity $\sigma_{xx}(\omega)$ simplifies to
\begin{eqnarray}\label{Op_B0}
\Re\big[\sigma_{xx}(\omega)\big] & = &  \frac{e^2}{4\pi \hbar} \frac{1}{\hbar \omega} 
\int_{0}^{\infty} \int_{0}^{2\pi} k^5 H(\theta) dk d\theta \nonumber\\
&\times &\big[ f(E_-) - f(E_+) \big] 
\delta \big(\hbar\omega - 2k^3 \Delta(\theta) \big) \nonumber\\
& = &\frac{e^2}{16\pi \hbar}\int_0^{2\pi}d\theta \frac{\sin ^2\theta 
\big[ 3 + \eta ^2-4\eta \sin 2\theta\big]^2}{3[1 + \eta^2 - 2 \eta \sin(2\theta) ]^2} 
\nonumber\\
& \times & \big[f(E_-(k_\omega))-f(E_+(k_\omega))\big],
\end{eqnarray}
with $k_{\omega}^3 = \hbar \omega/2\Delta(\theta)$ and $\eta=\beta/\alpha$.

{\bf Pure Rashba ($\beta=0$):} In the absence of DSOI ($\beta=0$), 
the absorptive part of the optical conductivity at finite temperature is given by
\begin{eqnarray}\label{Op_B0R}
\Re[\sigma_{xx}(\omega)] & = &\frac{3e^2}{16 \hbar} 
\big[f(E_-(k_\omega))-f(E_+(k_\omega))\big],
\end{eqnarray}
with $ k_{\omega}^3 = \hbar \omega/2\alpha $.
At zero temperature we have
\bearr
\Re [\sigma_{xx}(\omega)] & = &\frac{3e^2}{16 \hbar}
\big[ \Theta(E_-(k_\omega) - \mu)-\Theta(E_+ (k_\omega) - \mu) \big] \nn,
\eearr
where $\Theta(x) $ is the usual unit step function.

Depending on the carrier density $(n_h)$ and spin-orbit coupling strength ($\alpha$), 
there must be an upper and lower limits of the photon energy ($E_p = \hbar \omega$)
in order to have transitions from the initial state $\lambda = -1$ to
the final state $\lambda = +1 $. 
We use the following parameters for various plots: charge carrier density 
$ n_h = 2.4 \times 10^{15}$ m$^{-2}$ and
heavy hole mass $m=0.41 m_0$ with $m_0 $ is the bare electron mass.
In Fig. 3, we plot the optical conductivity $\sigma_{xx}(\omega)$ vs photon energy
$E_p $ for fixed $\alpha = 0.1$ eV nm$^{3}$ at four different temperatures. 
At $T=0$, the interband transitions take place only when photon energy 
satisfies the following inequality: 
$ 2 \alpha (k_{f}^{0,-})^3 \leq \hbar \omega \leq 2 \alpha (k_{f}^{0,+})^3 $
and the optical conductivity becomes box function with the edges at 
$ E_{\rm edge}^{\pm} = 2 \alpha (k_f^{0,\pm})^3$.
The width of the optical absorption is then
$ \Delta_b = 2 \alpha [(k_f^{0,+})^3 - (k_f^{0,-})^3] $, whose variation with
$n_h $ and $\alpha $ are shown in Fig. 4.
At finite temperature, the optical conductivity deviates from the box function
and smears beyond the box edges.
Moreover, the conductivity at the box edges 
($E_{\rm edge}^{\pm})$ is always $\sigma_0/2$ because of 
the nature of the Fermi distribution function. 
The peaks in the optical conductivity at finite $T$ is located
near the center of the box and it is given by 
$ E_{\rm peak} \simeq  (E_{\rm edge}^+ + E_{\rm edge}^-)/2 
= \alpha [(k_f^{0,+})^3 + (k_f^{0,-})^3] $.
Figure 4 shows that $\Delta_b$ increases
with $n_h $ and $\alpha$.
We mention here that similar analysis can be done for the opposite case i.e. 
$\alpha =0 $ but $\beta \neq 0 $. 
The optical conductivity at zero temperature will be 
$ \sigma(\omega) = e^2/(48 \hbar) $ which is 9 times less than $\beta=0 $ case.
Other results will be the same as for $\alpha \neq 0 $ but $\beta =0 $ case.

\begin{figure}[!htbp]
\begin{center}\leavevmode
\includegraphics[width=95mm,height=65mm]{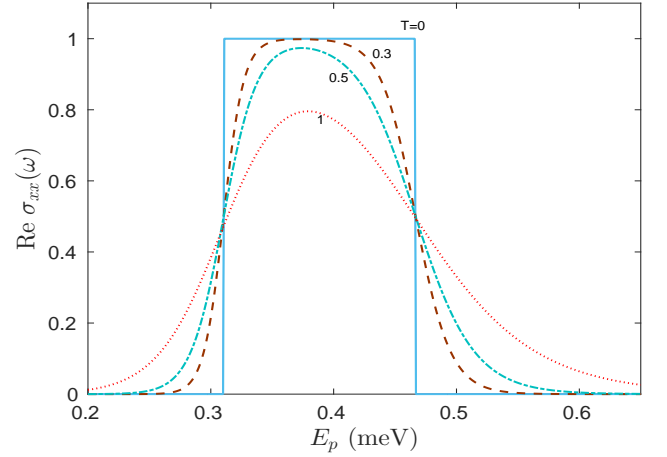}
\caption{(color online) Plots of $\sigma(\omega)$ 
(in units of $\sigma_0=3e^2/16 \hbar$) 
vs $E_p $ at four different temperatures.}
\label{Fig1}
\end{center}
\end{figure}

\begin{figure}[!htbp]
\begin{center}\leavevmode
\includegraphics[width=91mm,height=65mm]{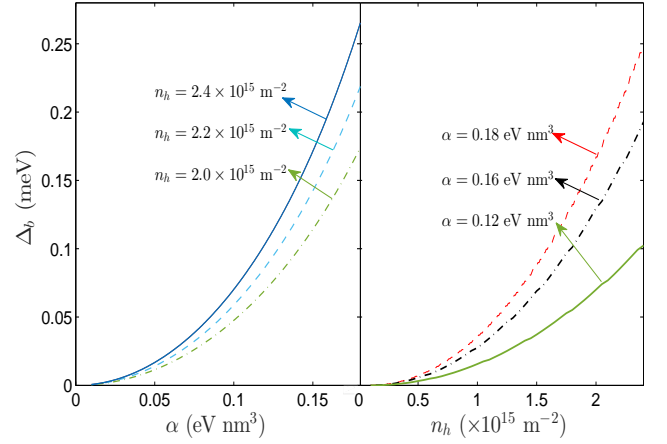}
\caption{(color online) (a) Bandwidth $\Delta_b$ vs $\alpha$ at various hole 
density and (b) $\Delta_b$ vs $n_h$ for various $\alpha$.}
\label{Fig1}
\end{center}
\end{figure}

{\bf Non-zero Rashba and Dresselhaus SOIs:} 
Now we discuss how the simultaneous presence of both the spin-orbit interactions  
modifies the behavior of the optical conductivity.
Similar to the previous case, the optical transitions between the initial 
state $ \lambda = - 1 $
and the final state $ \lambda = + 1$ can take place only when photon energy
satisfies the following inequality:
$ \epsilon_{-}(\theta) \leq \hbar \omega \leq \epsilon_{+}(\theta) $
with $ \epsilon_{\pm} = 2 (k_{f}^{\pm}(\theta))^3  \Delta(\theta) $. 
The values  of $ k_{f}^\pm(\theta) $ are the numerical solutions of
the two cubic equations
$ (\hbar k_{f}^\pm)^2/2m^* \pm (k_{f}^\pm)^3 \Delta(\theta) - E_f = 0 $,
where $E_f $ is the Fermi energy for given values of $\alpha$, $\beta$ and $n_h$.
\begin{figure}[!htbp]
\begin{center}\leavevmode
\includegraphics[width=167mm,height=135mm]{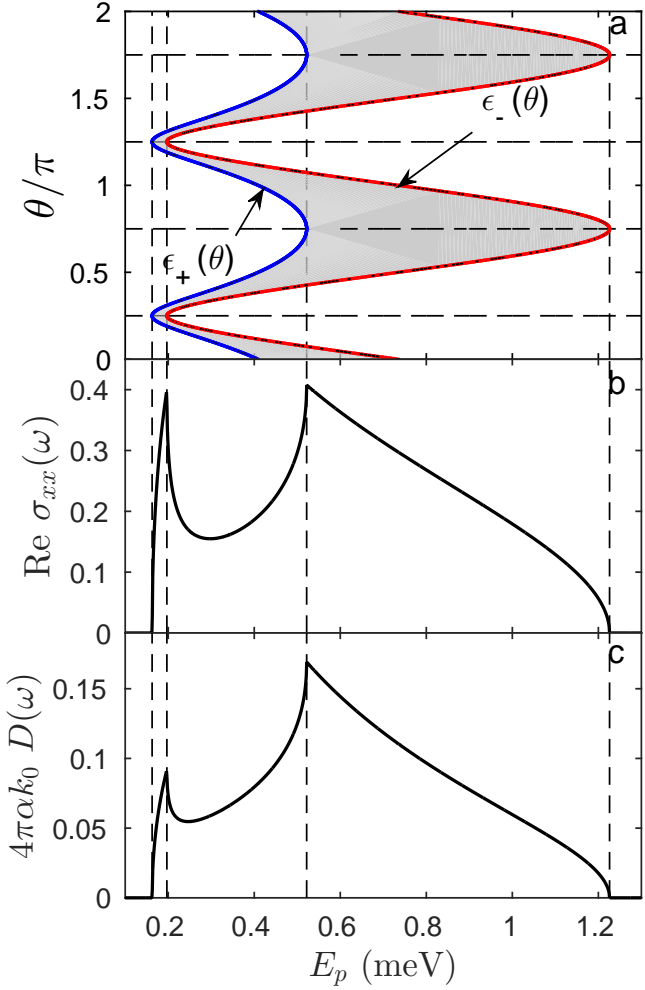}
\caption{(color online) Top panel: 
$ \epsilon_{\pm} = 2(k_{f}^{\pm}(\theta))^3 \Delta(\theta) $, 
Middle panel: $ \sigma_{xx}(\omega) $ (in units of $\sigma_0$) vs $E_p$ and 
bottom panel: joint density of states $D(\omega)$ with $k_0=\sqrt{2\pi n_h}$. 
Here $\alpha=0.12$ eV nm$^3$, $\eta=0.6$.}
\label{Fig1}
\end{center}
\end{figure}

In the top panel of Fig. 5, $\epsilon_{\pm}(\theta) $ vs photon energy 
are plotted. The shaded angular region contribute to the optical transitions.
The interband optical conductivity $\sigma_{xx}(\omega)$ vs $E_p$ is displayed 
in the middle panel of Fig. 5. 
We see that the optical transition begins and ends at $E_p = 0.162$ meV and 
$E_p = 1.226 $ meV, respectively. 
Looking at the top panel, one can see that these values correspond to 
$ \epsilon_+(\pi/4) = \epsilon_+(5\pi/4) =\epsilon_1$
and $ \epsilon_-(3\pi/4) = \epsilon_-(7\pi/4) = \epsilon_4$, respectively.
The minimum (maximum) photon energy $ \epsilon_1 (\epsilon_4)$ needed for interband 
optical transitions correspond to the excitation of a heavy hole with the Fermi wave vector
$ k_f^{+}(\theta) (k_f^{-}(\theta)) $ at $ \theta = \pi/4 $ or $ 5\pi/4 $ 
$ (\theta = 3\pi/4 $ or $ 7\pi/4 $).
The optical absorption edges of Fig.5 are exactly $ \epsilon_1$ and $\epsilon_4$.
Moreover, two peaks of the optical conductivity occur 
at $E_p = 0.196$ meV and $E_p = 0.522$ meV. 
It is easy to see from the top panel of Fig. 5 that these values correspond to 
$\epsilon_-(\pi/4) = \epsilon_-(5\pi/4)=\epsilon_2$ 
and $\epsilon_+(3\pi/4) = \epsilon_+(7\pi/4) =\epsilon_3$, respectively.

\begin{figure}[!htbp]
\begin{center}\leavevmode
\includegraphics[width=95mm,height=80mm]{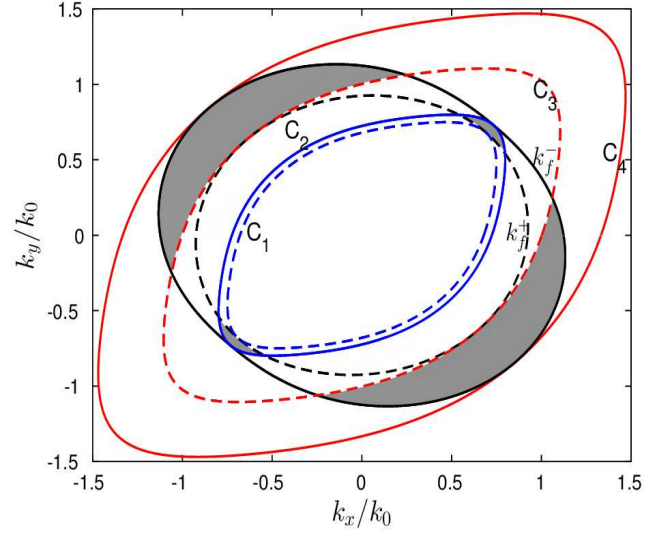}
\caption{(color online) This figure shows the Fermi contour 
${k_{f}^+}(\theta)$ (dotted-black), $k_{f}^-(\theta))$ (solid-black), 
the constant energy-difference 
$C(\hbar \omega) = \hbar\omega = 2\Delta(\theta){k}^3$ for 
$\hbar \omega = \epsilon_1$ (dashed-blue), 
$\hbar \omega = \epsilon_4$ (solid-blue), 
$\hbar \omega_2 = \epsilon_2$ (dashed-red) and 
$\hbar \omega_4 = \epsilon_3$ (solid-red).}
\label{Fig1}
\end{center}
\end{figure}  

In order to understand these behaviors we plot a constant 
energy-difference curve
$ E_{g}({\bf k})  = 2\Delta(\theta) k^3 = \hbar \omega $ 
for $ \hbar \omega = \epsilon_1$ ($C_1$: dotted-blue), 
$ \hbar \omega = \epsilon_2$ ($C_2$: solid-blue), $ \hbar \omega = \epsilon_3$ 
($C_3$: dashed-red) and $ \hbar \omega = \epsilon_4$ ($C_4$: solid-red) in Fig. 6. 
Because of the angular anisotropy in the dispersion relation, 
the optical conduction becomes ${\bf k}$-selective as shown in the shaded portions 
of Fig. 6 where the $C_i$'s ($i=1,2,3,4$) intersect with the two Fermi lines 
$ k_{f}^{+}(\theta) $ (dotted) and $ k_{f}^{-}(\theta) $ (solid), respectively.

Two peaks in the optical conductivity can be explained by analyzing the
joint density of states.
Usually, the absorptive part of the optical conductivity is
characterized by the joint density of states between the spin-split branches, 
which is given by
\begin{eqnarray*}\label{dosB0}
D(\omega)=\frac{1}{(2\pi)^2}\int d^2k \, [f(E_+)-f(E_-)]\delta(E_g({\bf k}) - \hbar \omega).
\end{eqnarray*}
Using the standard approach, we can write the joint DOS as
\begin{eqnarray}\label{dosB0}
D(\omega) & = & \frac{1}{(2\pi)^2}\int  
\frac{dC[f(E_+(k_\omega))-f(E_-(k_\omega))]}{|\partial_k  E_g({\bf k})  |_{E_g = \hbar \omega}},
\end{eqnarray}
where $dC $ is the line element along the contour and $k_\omega=(\hbar\omega/2\Delta(\theta))^{1/3}$.
The peaks appear in Fig. 5 whenever $ |{ \partial}_k E_g({\bf k}) | $ in the
joint DOS attains a minimum value. 
Therefore, the two peaks correspond to the van Hove singularities in the joint
density of states.
The first (second) peak is at a photon energy
$\hbar \omega_2 (\hbar \omega_3 $) for which the longer(shorter)
 axis of the curve $C_i $ coincides with the
Fermi line $k_{f}^{-}(\theta) (k_{f}^{+}(\theta) )$.
The joint density of states for finite $\eta$ is shown in the bottom panel
of Fig. 5. Locations of the two peaks as well as the optical absorption edges 
in the optical conductivity are exactly described by the joint density of states.
The asymmetric splitting at the Fermi level along the symmetry axis
$ k_y = \pm k_x$ is thus responsible for the peaks at $\epsilon_2$ and $\epsilon_3$,
respectively.
The magnitude and the non-symmetric shape of the optical conductivity is controlled
by the factor $ H(\theta)$.

For two-dimensional systems, van Hove singularities are classified into three
types based on the nature of change of the energy gap $E_g({\bf k})$ as we go away from
the singular points \cite{ssp-book}. 
This can be obtained by using the Taylor series expansion of $E_g({\bf k}) $ around the
singular points ${\bf k }_s $ at which the energy difference attains minimum value.
Here the singular points are at $ {\bf k}_s =(k,\pi/4 \textrm{ or } 5\pi/4)$ and
$ {\bf k}_s = (k,3\pi/4\textrm{ or } 7\pi/4)$.
Expanding $ E_g({\bf k}) $ around ${\bf k}_s$ as 
$ E_g({\bf k}) = E_g({\bf k}_s) + \sum_{i} b_i (k_i - k_{si})^2 $ with
$i =x,y$ and the expansion coefficients are
$2 b_i = \frac{\partial^2 E_g({\bf k})}{\partial k_{i}^2}|_{{\bf k}_s} $.
The classification of the van Hove singularities are based on how many coefficients
($b_i$) are negative. 
For the present system, the coefficients correspond to the expansion about 
the singular point $ {\bf k}_s =(k,\pi/4)$ are given by
\begin{eqnarray*}
b_x =\alpha k\frac{9(1-\eta)^2+12\eta}{2(1-\eta)},
\quad b_y =\alpha k\frac{9(1-\eta)^2-4\eta}{2(1-\eta)}\\
\end{eqnarray*}
and the coefficients correspond to the expansion about the singular 
point $ {\bf k}_s =(k,3\pi/4)$ are
\begin{eqnarray*}
b_x =\alpha k\frac{9(1+\eta)^2+4\eta}{2(1+\eta)},
\quad b_y =\alpha k\frac{9(1+\eta)^2-12\eta}{2(1+\eta)}.\\
\end{eqnarray*}

The type of singularities that arise are summarized in the table below:

\begin{table}[!htbp]
\begin{center}
\centering
\begin{tabular}{|M{1.2cm}|c|M{1cm}|M{2.6cm}|M{1.6cm}|} 
\hline
\hline
Singular point & $\eta=\beta/\alpha$ & $b_x$ & $b_y$ & Type of singularity\\\hline
\multirow{4}{*}{$(k,\pi/4)$} & \multirow{2}{*}{$\eta <1$} & \multirow{2}{*}{$>0$} 
& $>0$ for $\eta< \eta_l$ & $M_0$\\\cline{4-5}
& & & $<0$ for $\eta> \eta_l $ & $M_1$
 \\\cline{2-5}
& \multirow{2}{*}{$\eta >1$} & \multirow{2}{*}{$<0$} & $>0$ for $\eta<\eta_h $ & $M_1$\\\cline{4-5}
& & & $<0$ for $\eta> \eta_h $ & $M_2$\\
\hline
{$(k,3\pi/4)$} & -- & $ >0 $ & $ >0 $ & $ M_0 $\\ \hline
\end{tabular}
\caption{Table showing the type of singularity. 
Here, $\eta_l=(11-2\sqrt{10})/9$ and $\eta_h=(11+2\sqrt{10})/9$) are 
the solutions of the quadratic equation $9\eta^2-20\eta+9=0$.}
\end{center}
\end{table}

The optical conductivity versus $E_p$ at different values of $\eta $ at zero temperature
is shown in Fig. 7. 
Similarly, $\sigma(\omega) $ versus $E_p$ at different temperatures for a given value of
$\eta$ is shown in Fig. 8.

\begin{figure}[!htbp]
\begin{center}\leavevmode
\includegraphics[width=93mm,height=75mm]{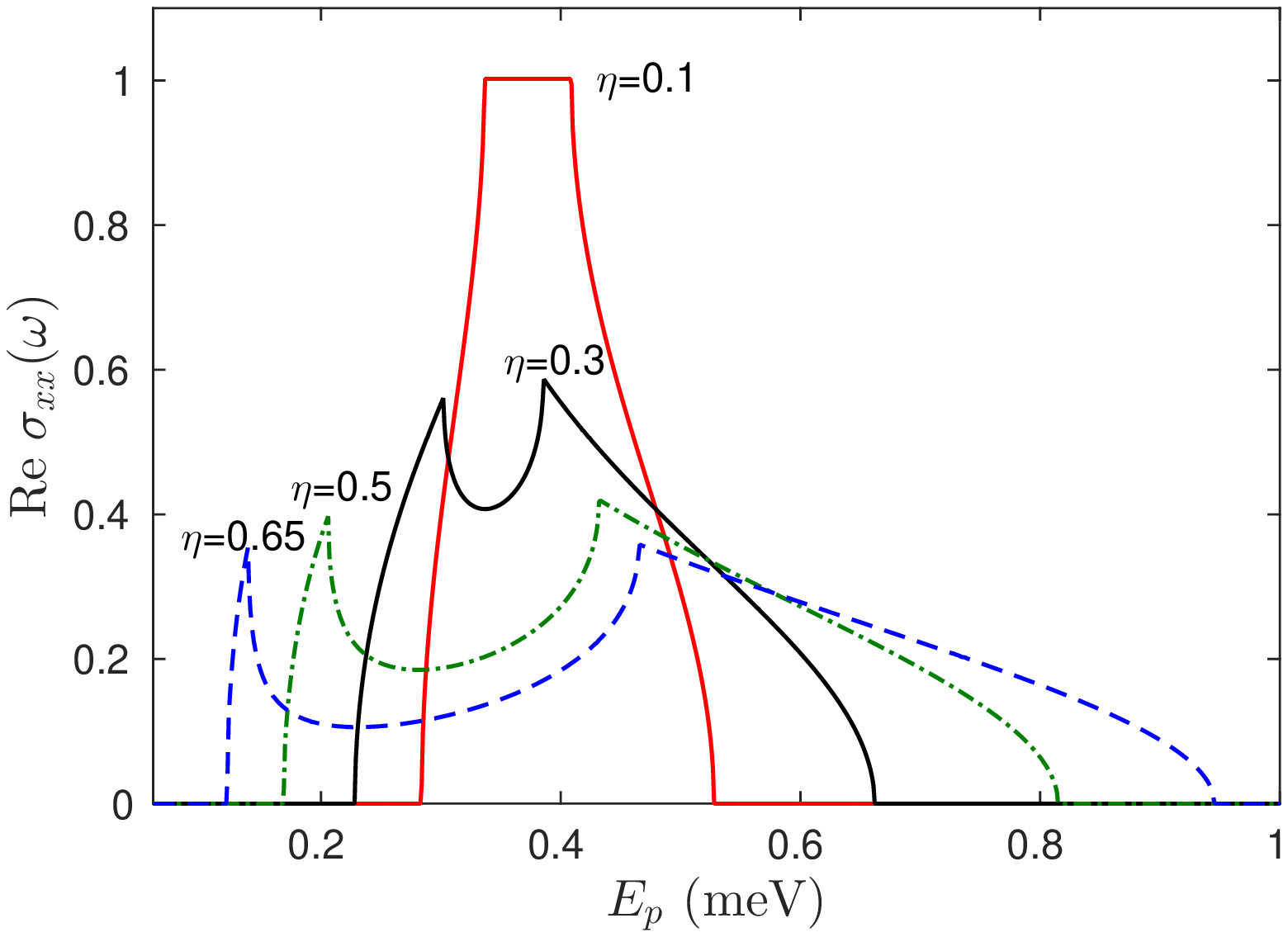}
\caption{(color online) Optical conductivity Re $\sigma_{xx}(\omega)$ in units of $\sigma_0$ 
for several values of $\eta$ with $\alpha=0.1$ eV nm$^3$.}
\label{Fig1}
\end{center}
\end{figure}

\begin{figure}[!htbp]
\begin{center}\leavevmode
\includegraphics[width=93mm,height=75mm]{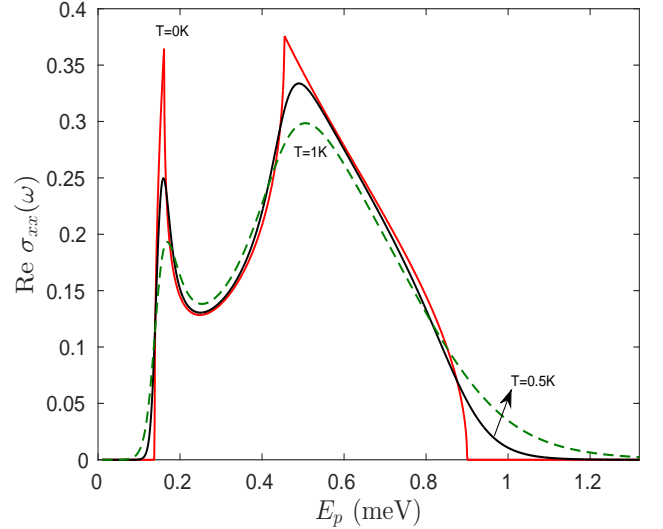}
\caption{(color online) Optical conductivity Re $\sigma_{xx}(\omega)$ in units 
of $\sigma_0$ for several values of temperature $T$ with $\alpha=0.1$ eV nm$^3$ and $\eta=0.6$.}
\label{Fig1}
\end{center}
\end{figure}

Now we shall point out here the main differences between the electron and heavy hole systems.
Unlike two-dimensional electron gas with $k$-linear RSOI and DSOI, the optical conductivity of two-dimensional hole gas
with $k$-cubic RSOI and DSOI does not vanish
for $\alpha = \beta $ case. This is related to non-zero Berry phase
of two-dimensional hole gas with equal strength of $k$-cubic RSOI and DSOI.
For realistic systems, the minimum photon energy needed to trigger the
optical transition in hole gas is one order of magnitude smaller
than that of electron gas.
In electron systems, heights of the two peaks are unequal, whereas they are more or less same
for hole gas at very low temperature.

\section{Summary and conclusion}
We have presented detailed analysis of zero-frequency Drude weight and optical
conductivity of a two-dimensional heavy-hole gas with $k$-cubic RSOI and DSOI, at both zero and non-zero temperature. 
We obtained an analytical expression of the Drude weight for Rashba interaction only.
It is shown that the Drude weight  deviates from the linear density dependence,
in contrast to the case of electron gas with and without $k$-linear spin-orbit interactions.
It decreases with the increase of the spin-orbit couplings.
We have identified a connection between the optical conductivity and
the Berry connection. 
On contray to the electron case, for equal strength of the Rashba and Dresselhaus couplings, 
the optical conductivity remains finite. This is due to fact that
Berry phase is not zero for equal strength of $k$-cubic spin-orbit couplings. 
The bandwidth increases with increase of the hole density as well as spin-orbit couplings.
It is seen that the minimum photon energy required to set in the optical transition
in hole gas is one order of magnitude smaller than that of electron gas.
We have classified the type of the two van Hove singular points.

\appendix
\setcounter{secnumdepth}{0}
\section{Appendix: Calculation of Drude weight for system with linear RSOI}
Here we consider two-dimensional electron gas with $k$-linear
spin-orbit interaction and calculate the Drude weight using Eq. (\ref{drude-weight}).
The Hamiltonian for this system is given by
\begin{eqnarray}
H_e = \frac{ p^2}{2m_e}+ \frac{\alpha_e}{\hbar}\big(\sigma_x {p}_y - \sigma_y {p}_x\big),
\end{eqnarray}
where $m_e$ is the electron's effective mass and $\alpha_e$ is the strength of the Rashba SOI.
The corresponding energy eigenvalues and eigenstates are 
$E_\lambda({\bf k})=\hbar^2 k^2/2m_e + \lambda\alpha_e k$ and 
$\psi^\lambda_{\bf k}({\bf r})=e^{i{\bf k} \cdot {\bf r}}\begin{pmatrix}
1, & -i\lambda e^{i\theta}
\end{pmatrix}^T/\sqrt{2\Omega}$, respectively. 
Here $\theta=\tan^{-1}(k_y/k_x)$ and $T$ denotes the transpose operation.
The Fermi energy ($E_f$) can be obtained as 
$E_f = \pi \hbar^2 n_e /m_e - m_e\alpha_{e}^2/\hbar^2$ with $n_e$ being the electron density.

The $x$-component of the velocity operator is 
$ \hat{v}_x = {p}_x/m_e - (\alpha_e/\hbar) {\sigma}_y $.
Its expectation value is 
$ \langle \hat{v}_x \rangle 
= (\hbar/m_e)(k + \lambda k_\alpha)\cos\theta$
with $k_\alpha=m_e\alpha_e/\hbar^2$.

Following Eq. (\ref{drude-weight}), we have the Drude weight as below
\begin{eqnarray*}
D_w^e & = & \frac{e^2}{4\pi m_e}\sum_\lambda \int d^2 {\bf k} (k+\lambda k_\alpha)^2 
\cos^2\theta\frac{\delta(k-k_{f}^\lambda)}{|k_{f}^\lambda+\lambda k_\alpha|}, 
\end{eqnarray*}
where $k_f^\lambda = -\lambda k_\alpha +\sqrt{2\pi n_e - k_\alpha^2}$  
are the spin-split Fermi wave-vectors.
The final expression of the Drude weight is now
\begin{eqnarray*}
D_w^e & = & \frac{\pi e^2}{m_e} \Big(n_e - \frac{m_e^2\alpha_e^2}{2\pi\hbar^4} \Big),
\end{eqnarray*}
which is the same as given in Eq. (44) of Ref. \citep{agarwal}.

\end{document}